\documentclass[preprint]{aa}
\usepackage{graphicx}

\begin{document}

\title{Probing the evolution of galaxy clusters with the SZ Effect}

\author{A. Cavaliere\inst{1,2} \and A. Lapi\inst{1,3,4,5}}
\institute{Dip. Fisica, Univ. `Tor Vergata', Via Ricerca Scientifica 1, 00133
Roma, Italy \and INAF, Osservatorio Astronomico di Roma, Via Frascati 33,
00040 Monteporzio, Italy \and SISSA, Via Bonomea 265, 34136 Trieste, Italy
\and INFN - Sezione di Trieste, Via Valerio 2, 34127 Trieste, Italy \and INAF
- Osservatorio Astronomico di Trieste, via Tiepolo 11, 34131 Trieste, Italy}

\date{\today}

\abstract{In galaxy clusters the thermal Sunyaev-Zel'dovich (SZ) effect from
the hot intracluster medium (ICM) provides a direct, self-contained measure
of the pressure integrated over crossing lines of sight, that is
intrinsically independent of redshift and well suited for evolutionary
studies. We show in detail how the size of the effect and its pattern on the
sky plane are directly related to the entropy levels in the ICM, and how they
characterize the cluster cores and outskirts independently. We find that at
redshifts $z\lesssim 0.3$ the signals to be expected in the cores
considerably exceed those detected at $10'$ resolution with the Planck
satellite. We propose that at 1' resolutions as implemented on recent ground
instrumentation for mapping features in individual clusters, the average
patterns of the SZ signals can provide a direct and effective way to find and
count cool, low-entropy cores and hot, high-entropy outskirts out to $z \sim
2$. Such counts will tell the timing and the mode of the processes that drive
the evolution of the ICM from the distant to the local cluster population.}

{\keywords{cosmic background radiation -- galaxies: clusters: general --
X-rays: galaxies: clusters -- methods: analytical.}

\authorrunning{Cavaliere, Lapi}
\titlerunning{Evolution of Galaxy Clusters from SZ Observations}

\maketitle

\section{Introduction}

The thermal SZ effect (Sunyaev \& Zel'dovich 1972, 1980) from a galaxy
cluster is produced when cold cosmic microwave background (CMB) photons are
upscattered with Thomson cross section $\sigma_T$ by the electrons in the hot
electron-proton plasma constituting the intracluster medium (ICM). The
resulting brightness is proportional to the Comptonization parameter
\begin{equation}
y(s) = {\sigma_T\over m_e c^2}\, \int\, {\rm d} \ell~ p_e,
\end{equation}
given by the electron thermal pressure $p_e$ integrated along the line of
sight (l.o.s.) $\ell$ crossing the cluster at the projected distance $s$ from
the center.

The effect carries a specific spectral signature, as the frequency-depending
prefactor to $y(s)$ goes from negative to positive through $217$ GHz.
Meanwhile, its $10^{-4}$ K amplitudes encase a wealth of information
concerning the ICM. The SZ effect provides a self-contained probe of the ICM
pressure, given by $p\simeq 1.9 \, p_e$ in the presence of thermal
equilibrium and cosmic abundances. By its linear nature such a probe is
insensitive to isobaric clumpiness; it is also intrinsically independent of
the redshift $z$, and thus well suited for evolutionary studies.

The SZ effect is now providing comprehensive catalogues that comprise
hundreds of clusters selected out to $z \sim 1$. So far, such selections have
been mostly joined to X-ray analyses to describe the pressure profiles of the
ICM through the cluster outskirts and within the cores at radii $r_c \lesssim
0.2 \, R_{500}$\footnote{The radius $R_{500}$ embraces a mean overdensity
$500$ times above the critical density at the epoch of cluster formation, and
is related by $R_{500} \simeq R/2$ to the virial radius $R$. We adopt the
flat, currently accelerating $\Lambda$CDM cosmology (see Hinshaw et al. 2013,
\textsl{Planck} Collaboration XVI 2014) with round parameters: Hubble
constant $H_0 = 70$ km s$^{-1}$ Mpc$^{-1}$, matter density $\Omega_M = 0.3$
including a baryon density $\Omega_b = 0.04$; the Hubble function reads $h(z)
= [\Omega_M \, (1+z)^3 + \Omega_{\Lambda}]^{1/2}$.}. For example, McDonald et
al. (2013, 2014) followed up with \textsl{Chandra} 80 clusters selected out
to $z \simeq 1.2$ by the \textsl{South Pole Telescope} with SZ effects at
high signal-to-noise $S/N > 5$ (see Reichardt et al. 2013). They combined
highly resolved profiles of the number density $n$ with temperatures $T$
averaged within six ranges of $z$ so as to gather sufficient counts for X-ray
spectroscopy and to outline profiles of $p \propto n\, T$. The focus was on
profiles with a central temperature dip defining cool core clusters, as
opposed to the extended plateau marking non-cool core clusters (CCs and NCCs;
see Molendi \& Pizzolato 2001, Hudson et al. 2010).

On the other hand, the \textsl{Planck} Collaboration V (2013) analyzed their
SZ data concerning $62$ clusters mostly at $z\lesssim 0.3$, and obtained the
cluster signals through the procedure proposed and discussed by Melin et al.
(2006). This included several steps: sky maps at several frequencies were
cleaned of contaminating sources and deconvolved from the instrumental beam
to high $S/N$, and patterns $y(s)$ were extracted and finally deprojected to
obtain radial pressure profiles $p(r)$. The \textsl{Planck} resolution around
$10'$ can retrieve direct information concerning such profiles only at radii
$r\ga 0.5\; R_{500}$, depending in detail on $z$. To reach inner regions, the
observed profiles were matched to \textsl{XMM-Newton} X-ray information
resolved on $1'$ scales, with the help of fitting formulae of the generalized
NFW kind that combine inverse power laws of $r$ (Nagai et al. 2007, Arnaud et
al. 2010, Bonamente et al. 2012).

Such a template is attractively simple. By the same token, it is constrained
to the $z$ dependence given by $P_{500} \propto h(z)\,^{8/3}$, and to
centrally diverging shapes $p(r) \propto r^{-0.3}$. In the outskirts, the
pressure would always decline faster than $p(r)\propto r^{-3.5}$; there the
signals would rapidly weaken, and noise combines with scatter introduced by
deprojection to yield increasing uncertainties that blur physical
information. Simplicity is paid for with wide dispersions.

Profiles and fits are best addressed in terms of the specific entropy
\begin{equation}
k \equiv p/n^{5/3},
\end{equation}
the crucial quantity that controls the ICM sinking or rising (see Bower 1997;
Voit 2005) in the gravitational potential well dominantly provided by the
dark matter (DM) halo. We shall see how entropy also yields more and closer
information concerning the physical processes that prevail at the inner and
outer ends; their understanding is urged by everybody in the field, including
\textsl{Planck} Collaboration V (2013) and XVI (2014), Eckert et al. (2013),
McDonald et al. (2014).

On the other hand, hundreds of SZ-detected clusters are still available, many
with measured values of $z\ga 0.3$, from \textsl{Planck} (see \textsl{Planck}
Collaboration XXIX 2013) and from ground-based surveys (see Carlstrom et al.
2011, Hasselfield et al. 2013, Reichardt et al. 2013). Below we propose a
fast way to scan and search such a large data body directly in terms of
$y(s)$ on arcmin scales, to retrieve information concerning the ICM evolution
and its driving processes at increasing $z$.

Our aim warrants a detailed analysis as given in the next sections along the
following layout. In Sect. 2 we make contact with previous work in showing
how entropy governs the pressure profiles in the ICM. In Sect. 3 we detail
how their inner and outer shapes independently relate to the entropy levels
there. In Sect. 4 we propose and discuss our main point, centered on how such
levels can be directly and separately probed with the use of simple scaling
features marking the outer and inner SZ signals. In Sect. 5 we discuss how
such features depend on the chemical composition of the ICM, on non-thermal
contributions to pressure, and on the presence of multiple phases. Finally,
Sect. 6 concludes with the specific information to be expected on the
physical processes that shape the ICM.

\section{Advanced pressure profiles}

Our approach starts from the entropy reference run $k(r)=k_c+k_R\,(r/R)^a$
modulated by two intrinsic and \emph{independent} parameters, namely, the
core value $k_c$ and the slope $a$ of the outer, rising ramp (see Voit 2005,
Lapi et al. 2005, Cavagnolo et al. 2009). Both originate from gravitational
processes during the closely self-similar formation process of the cluster DM
halos (Zhao et al. 2003, Wang et al. 2011, Lapi \& Cavaliere 2009, 2011).

In the ICM the initial entropy levels are expected to be high. Values $k_c
\sim 10^2$ keV cm$^2$ are set at $z_t \sim 1 - 2$ by the first collapse and
virialization of the cores from an initial density perturbation, as the gas
mass density $n\,m_p$ is raised by some $10^2$ and the temperature to
$G\,M(<r_c)\,m_p/10\,r_c\sim$ a few keVs (see Kravtsov \& Borgani 2012, and
Cavaliere \& Lapi 2013, hereafter CL13). Outer ramp slopes $a \simeq 1$
(Tozzi \& Norman 2001) are built up during the inside-out cluster growth fed
over several Gyrs by inflows of outer DM and gas through the virial boundary.
The DM goes to increase the halo mass and the concentration $C \simeq 3.5\,
(1+z_t)/(1+z)$ (e.g., Zhao et al. 2003, Prada et al. 2012). The gas drives at
$r\sim R$ a layer of accretion shocks with Mach numbers given by
$\mathcal{M}^2\equiv G M(<R)/R\,c_s^2\ga 3$ over the outer sound speed $c_s$;
the shocks promptly thermalize and stratify a share exceeding $60\%$ of the
gas infall energy (see CL13; also Vazza et al. 2009, Kravtsov \& Borgani
2012). These processes concur to set an initial state of \emph{high} entropy
(HE) throughout the ICM.

In the dense cores, however, the ICM radiates strongly in X-rays by optically
thin thermal bremsstrahlung. Thus over times $t_c \approx 0.3\,
(k_c/10~\mathrm{keV~ cm}^2)^{1.2}$ Gyrs the ICM is expected to undergo
accelerated cooling, and to gently settle down to a central state of lower
entropy with $k_c \simeq$ a few keV cm$^2$, until losses are roughly balanced
by gains from bubbling or rekindled AGNs in the central member galaxies (see
Fabian 2012). On the other hand, on comparable timescales (depending in
detail on cluster age and environment richness, see \S~6 and CL13) outer
inflows are expected to subside and drive diminishing entropy production, so
that the outward ramp flattens out to slopes $a \lesssim 0.7$. Thus
conditions of \emph{low} entropy (LE) tend to eventually prevail throughout
the ICM; in particular, synchronous core cooling and diminishing outer
entropy production would smoothly take the whole ICM from HE to LE overall
conditions, and provide the simple picture we shall entertain first.

The actual conditions may be richer, however, and intermediate types will be
discussed in \S~6. For example, initially many small mergers with their bound
gas may keep the outer ICM cool; later massive mergers may hit and reheat the
inner ICM for a while (see McCarthy et al. 2007), but pressure (when not
sustained by dynamical stresses like that prevail in shocks) recovers from
such perturbations over a few sound crossing times, so it is convenient to
focus first on the condition for thermal, hydrostatic, and spherical
equilibrium in the form
\begin{equation}
{{\rm d}p\over {\rm d}r} = - ~ n\,m_p\, {G M(< r)\over r^2}~.
\end{equation}
This is governed by the specific gravitational force $G\,M(<r)/r^2$ produced
by the mass distribution in the DM halo (see Cavaliere et al. 2009, hereafter
CLFF09) acting on the ICM mass density $n\, m_p$.

The entropy $k$ as defined in \S~1 leads to expressing $n \propto
(p/k)^{3/5}$. With that, Eq.~(3) converts to an equality for ${\rm
d}p^{2/5}/{\rm d}r$; the equality straightforwardly integrates to yield the
pressure profile $p(r)$ joining onto the value $p_R/p_1\simeq
\mathcal{M}^2/3$, with the jump set by the boundary shocks above the outer
value $p_1$. The result writes
\begin{equation}
p(r)=\left[p_R^{2/5}+{2\over 5}\,\int_r^R{\mathrm d}x~{m_p\,G\, M(<x)\over
x^2\, k^{3/5}(x)}~\right]^{5/2}~,
\end{equation}
and shows how an entropy run $k(r)$ provides the spine to the corresponding
pressure profile $p(r)$. Specifically, our reference run $k(r)$ given at the
beginning of this section yields the examples of $p(r)$ illustrated in Fig.~1
for HE and for LE conditions throughout the ICM.

\section{Scaling laws}

From Fig.~1 it may be also realized that the core profiles depend strongly on
the central level $k_c$ while the outskirts are modulated by the outer slope
$a$, with an \emph{independent} role played by these parameters. For any
specific cluster their values can be pinned down (using the Supermodel
approach, see CLFF09) by choosing the parameters $k_c$ and $a$ so that the
model $p(r)$ fits the observed shapes. The procedure is detailed on the site
\texttt{http://people.sissa.it/$\sim$lapi/Supermodel/}.

Here we stress that \emph{before} (or even instead of) fitting detailed
pressure profiles, the HE/LE classification can be established simply by
looking at scaling features of $p(r)$ in terms of entropy. In the
\emph{cores} these read
\begin{equation}
p_c \propto k_c^{-5/8}~, ~~~~~{\rm and}~~~~\left[{\rm d}p/{\rm
d}r\right]_{{r^*}} \propto k_c^{-1} \propto p_c^{8/5}~,
\end{equation}
at radii $r^*\lesssim 0.1\, R_{500}$ (see CL13); the absolute logarithmic
slopes scale as
\begin{equation}
\left[{{\rm d} \ln p \over {\rm d} \ln r}\right]_{{r^*}} \propto k_c^{-1/3}
\propto p_c^{1/2}~,
\end{equation}
yielding an indicator akin to the cuspiness defined for the density profiles
by Vikhlinin et al. (2007). We note that the gradient ${\rm d}p/{\rm d}r$ in
the cores is proportional to the force $G M(<r)/r^2$ entering Eq.~(3). Toward
the center this vanishes for matter densities increasing less than $r^{-1}$
(see data and discussion in Newman et al. 2013) and for a finite ICM entropy,
as to yield a maximum of $p(r)$. Then the value of $p_c$ does not depend
strongly on the detailed choice of $r_\star$, nor on the presence of any
central bright galaxy.

Thus we see how neatly steep negative gradients in the cores \emph{correlate}
with high central levels $p_c$; both quantities are much larger for LEs than
for HEs, and combine into a clear mark of the former condition. This means
that, as $k_c$ lowers by a factor of 10, the central peak grows by a factor
of $4$, and the logarithmic slopes steepen by $2$.  Meanwhile, for $k_c\la
20$ keV cm$^2$ the steeply declining pressure $p(r)$ combines with the
briskly rising entropy $k(r)$ to yield in the temperature profile
$T(r)\propto p^{2/5}(r)\,k^{3/5}(r)$  a central dip  and a maximum at $r\sim
0.2\, R_{500}$, the features that mark the CC clusters as observed in X rays
(see CLFF09).

On the other hand, CLFF09 matched the boundary shock jumps to the adjoining
hydrostatic equilibrium in the \emph{outskirts}, and derived the scaling
$p(r)/p_R \propto (r/R)^{2a - 5}$. This yields
\begin{equation}
p(r) \propto r^{-2.8} ~, ~~~~~~{\rm and} ~~~~ p(r) \propto r ^{-3.6} ~~~ {\rm or} ~~
{\rm steeper}
\end{equation}
for fully developed HE conditions with $a\sim 1.1$ and for aged LE conditions
with $a \lesssim 0.7$, respectively. How well all such features apply to HE
and LE types can be checked in Fig.~1.

\begin{figure}
\centering
\includegraphics[width=\columnwidth]{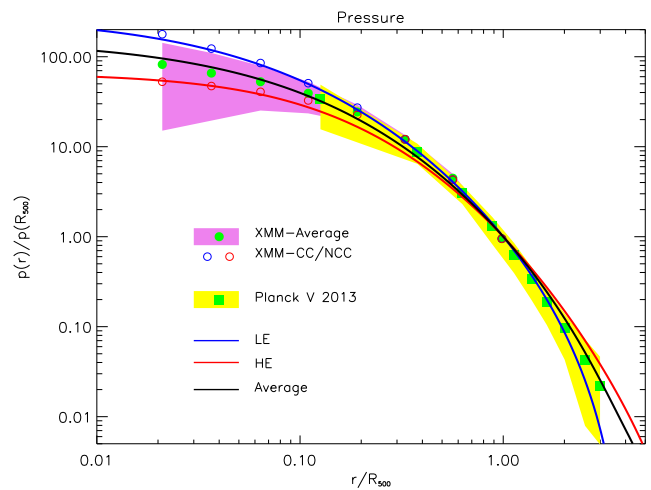}
\caption{Pressure profiles we expect using Eq.~(4) for a cluster in overall
HE conditions with central entropy level $k_c=100$ keV cm$^2$ and outer
entropy slope $a = 1.1$ (red line), and for a cluster in overall LE
conditions with $k_c=10$ keV cm$^2$ and $a = 0.7$ (blue line); the outcome
for a mixture of the two HE and LE templates in the $1:3$ proportion (black
line) is also presented. These profiles are compared with the SZ data from
\textsl{Planck}, and with the X-ray observations from \textsl{XMM-Newton}
(green squares, with the yellow shaded area illustrating the variance). All
data are taken from \textsl{Planck} Collaboration V (2013), and are
reproduced for the whole cluster sample (green circles with purple shaded
area) and for the CC and NCC subsamples (open blue and red circles,
respectively).}
\end{figure}

\section{Direct use of SZ patterns}

We now proceed toward our main aim, which is to see how \emph{direct} use can
be made of the SZ patterns $y(s)$ on the sky plane; these constitute the
primary observables, while the radial pressure profiles $p(r)$ require
delicate deprojections along the line of sight $\ell = (r^2 - s^2)^{1/2}$. In
Fig.~2 we plot the patterns $y(s)$ that we expect for a fully LE or HE
cluster, corresponding to the profiles of $p(r)$ given in Fig.~1.

In the \emph{cores}, the pattern peaks are given by $y_c = - 2 ~ \int_0^R
{\rm d}r ~r ~ {\rm d}p/{\rm d}r$, once Eq.~(1) is expressed in terms of $r$
and integrated by parts. When the central pressure peak is as high as applies
to LE conditions, its scaling in terms of central entropy dominates the
result although in a mitigated fashion, to yield
\begin{equation}
y_c \propto k_c^{- 5/8 + 0.1}.
\end{equation}

This is derived from the approximation given by CLFF09 for $p(r)$ in the
cores that reads
\begin{equation}
p(r) \simeq I^{5/2}(r) = [A_0\, e^{- A_1\,r^{A_2}}]^{5/2}~.
\end{equation}
The coefficient $A_0\propto k_c^{-1/4}$ scales up with decreasing $k_c$, as
illustrated by the steep dotted line in Fig.~2 (bottom panel) of CLFF09.
Meanwhile, $A_1$ and $A_2$ scale down slowly\footnote{Fiducial values for the
triple $(A_0, A_1, A_2)$ may be read out from Table 1 in CLFF09: a typical HE
cluster corresponds to $(8.5,2.2,0.8)$; a typical LE to $(17,2.3,0.65)$.},
and contribute to the milder rise for small $k_c$ shown by the solid line in
the same figure. In the present context, $A_0^{5/2}$ yields just the main
scaling $p_c \propto k_c^{-5/8}$ highlighted in Eq.~(5); it also dominates
$y_c$. To a close approximation, the derivative ${\rm d}p/{\rm d}r$ (see the
integration by parts spelled out above) implies that the factor $A_0$ is
multiplied by $A_1\, A_2$, which in turn mitigate the scaling as is shown in
Eq.~(8) and is borne out by comparing the peak ratios in Fig.~2 (dashed
lines) with those in Fig.~1.

\begin{figure}
\centering
\includegraphics[width=\columnwidth]{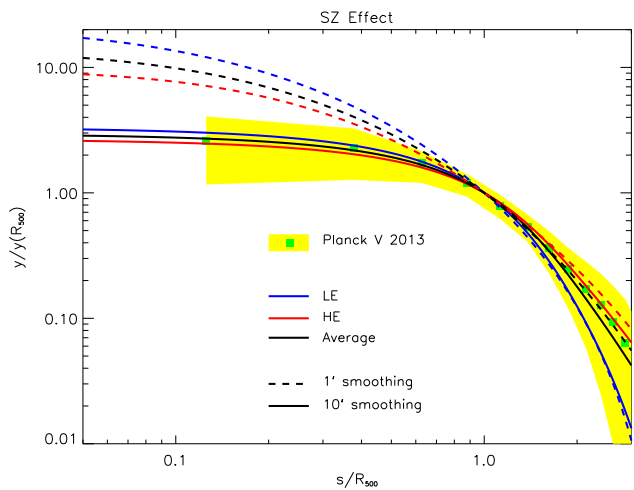}
\caption{Our predicted SZ patterns, corresponding to the
pressure profiles marked with the same colors as in Fig.~1. The dashed lines refer
to a resolution around $1'$ (closely matching the unconvolved patterns), and
the solid lines to a resolution around $10'$ like \textsl{Planck}'s. The data
are from SZ observations after \textsl{Planck} Collaboration V (2013; green
squares, with yellow shaded area showing the associated variance).}
\end{figure}

For a l.o.s. beyond the core, the integration in the form $y(s) = 2\,
\int_0^{\ell_{m}} {\rm d}\ell \, p[r(\ell)]$ (with $\ell^2_{m} = R^2 - s^2$)
shows how the runs of $p(r)$ are smoothed out to yield the slow (for HE), or
the mild (for LE) pattern decline shown in Fig.~2.

In the \emph{outskirts} the pressure profiles $p(r)\propto r^{2a - 5}$ with
slopes ranging from $a \simeq 1.1$ to $a \lesssim 0.7$ (see Eq.~7) correspond
to steeper and steeper declines of the patterns $y(s)$, as borne out by
Fig.~2 (from red to blue lines); at $s = 1$ ($r = R$) all plunge into the
boundary value $p_R$ set by virial shocks. Specifically, we obtain
\begin{equation}
y(s) \propto (1-s^2)^{1/2}/s^2, ~~ ~~~ {\rm and} ~~~ y \propto
(1-s^2)^{1/2}/s^4
\end{equation}
for $a =1$ (HE condition) and for $a = 0$ (low entropy condition),
respectively. In brief, by l.o.s. integration the profiles $p \propto r^{2a
-5}$ are smoothed out into the patterns $y(s) \propto s^{2a - 4}$, before a
final plunge at $r \simeq R$.

Thus the patterns corresponding to HE and LE conditions can be identified
from their specific inner and outer scaling laws given above, and may be
tested on stacked SZ observations as illustrated in Fig.~2. The solid lines
show the representation provided by our patterns for the \textsl{Planck} data
with their resolution around $10'$. The dashed lines show the considerably
\emph{larger} signals to be expected at a resolution of $1'$; they closely
approach the unconvolved patterns, and vary with entropy just as expected
from the scaling laws in Eqs.~(8) and (10). A hint at the plunging behavior
discussed above may be already discerned in the outer data (see green squares
and solid red/black lines in Fig.~2).

We propose that individual patterns $y(s)$ observed at resolutions
substantially \textit{better than $10'$} in the cores and in the outskirts be
\emph{directly} used for purposes of cluster classification and counts in
terms of entropy, dispensing with deconvolutions and attendant uncertainties.
We have shown that the classification can be pursued in terms of
\emph{independent} entropy levels in the cores and in the outskirts.

In a nutshell, we propose to \emph{directly} relate the patterns $y(s)$
observed with the SZ effect to entropy levels $k(r)$, so as to produce
independent and robust entropy maps in the ICM of the cores and of the
outskirts.

\section{Stability}

How do the above patterns respond to variance in equilibrium conditions and
chemical composition? Lapi et al. (2012a; see also references therein) have
shown that the total pressure $p$ supporting the equilibrium described by
Eq.~(3) may include -- especially in the outer region of LEs -- not only the
thermal pressure $p$, but also a non-thermal, turbulent contribution
$\delta~p$, to yield
\begin{equation}
p_{\rm tot} = (1 + \delta)~p
\end{equation}
with $\delta\lesssim 40\%$ slowly decaying inward of the shock layer. Such
conditions imply correspondingly lower levels of the thermal component $p(r)$
and of the related SZ signal $y(s)$; so the outer shapes discussed in \S~4
constitute effective \emph{upper} bounds. Likewise, $p_e$ is somewhat
decreased (relative to the thermal equilibrium value given in \S~1) for
electron temperatures still catching up with ions just downstream from the
boundary shocks (see Wong \& Sarazin 2009).

McDonald et al. (2014) show in their Fig.~6 that even for NCC clusters like
SPT-CLJ0014-4952 where X-ray imaging emphasizes scars in the ICM from mergers
and central features from sloshing, the SZ mapping is better at rendering
centroid location and overall circularity on the sky plane. Along the l.o.s.
the SZ signals will be anti-biased when ellipsoidal clusters are observed
parallel to their minor axis; for the cluster RX J1347.5-1145 Plagge et al.
(2013) find evidence of considerable compression, in line with Bonamente et
al. (2012). From statistics of spherically averaged models, Buote \& Humphrey
(2012) find mean biases $\sim 1\%$ with scatter $\la 10\%$, and less in the
cores. So our LE characterization will be minimally affected, while HE
outskirts will still provide valuable lower bounds to patterns $y(s)$.

On the other hand, at the center of relaxed LEs some Helium sedimentation may
occur (see discussions by Ettori \& Fabian 2006 and Bulbul et al. 2011); this
will decrease the Hydrogen abundance $X$ and increase the values of
\begin{equation}
p_e = (2 + 2 X)/(3 + 5X)~p \gtrsim 0.52\, p~.
\end{equation}
So $y$ will be enhanced, at least, in the core, though conceivably by less
than the $4 \%$ level given by doubling the cosmic He abundance. On this
account, the central values discussed in \S~4 constitute effective
\emph{lower} bounds. Thus the above considerations concur to widen the
\emph{difference} between LE and HE conditions both in the cores and in the
outskirts.

In a more radical vein, in LE cores the total entropy $k_c$ may be lowered to
a few keV cm$^2$ in low-$z$ clusters by the settling of a cooled phase, as
highlighted by finely resolved measurements of X-ray brightness (see
Panagoulia et al. 2014). This condition will make the central arcmin peaks of
$y(s)$ even higher, and easier to recognize with the SZ effect than indicated
by Eq.~(8).

\section{Discussion and conclusions}

The SZ effect from galaxy clusters can directly probe the thermal pressure
$p$ in their ICM out to redshifts $z\sim 1$, where the X-ray brightness is so
faint as to hinder measuring $T$ and obtain $p$ from $p\propto n\,T$. We have
shown in \S~3 how the pressure profiles $p(r)$ \emph{differ} in terms of
entropy, the basic quantity that provides the key to characterizing the ICM
thermal states in terms of low (LE) or high (HE) entropy conditions. To
distinguish LEs from HEs it is enough to focus on a few basic profile
features in the \emph{core} (peak rise correlated with height) and in the
\emph{outskirts} (steep vs. shallow decline) as detailed in \S~3.

Even more directly, we have highlighted in \S~4 how such a cluster
characterization can be carried out directly in terms of the \emph{primary}
SZ patterns $y(s)$, focusing again on their \emph{core} scaling and
\emph{outer} slopes. The task requires retrieving at resolutions $\sim 1'$
the considerable levels of SZ signals that went undetected in cores at
resolutions $\sim 10'$. Thus efficient use can be made of larger samples,
while deprojection uncertainties and template peculiarities are sidestepped.
Then probing the ICM with the SZ effect can provide direct and fast
statistical tests for redshift \emph{evolution} of the ICM in cores and in
outskirts.

The results may be compared with the simple picture anticipated in \S~2,
where radiative cooling of cores and diminishing gravitational heating of
outskirts $-$ though independent $-$ tend to proceed on comparable timescales
of several Gyrs. With growing lifetime the DM halo concentration $C(z)$
increases, while radiative cooling and diminishing heating concur to drive
the whole ICM away from its initial HE conditions marked by hot cores at $k_c
\sim 10^2$ keV cm$^2$ and steep outer slopes $a\simeq 1$. Eventually, the ICM
(coupled to its ambient by mass inflow and radiation outflow) would proceed
toward overall LE states of \emph{lower} entropy, with $k_c \lesssim 10 $ keV
cm$^2$ and $a \lesssim 0.7$.

At $z\lesssim 0.3$ we expect that our LE core conditions largely superpose to
the strong CC clusters (Hudson et al. 2010), as the HEs do with NCC ones. For
increasing $z$, we expect the statistical LE/HE ratio to drift \emph{below}
the values around $1/2$ prevailing locally (see Rossetti \& Molendi 2010,
Santos et al. 2010). We show in Figs.~1 and 2 and their captions a ratio
LE/HE $\lesssim 1/3$ to be more fitting for the SZ-selected sample in
\textsl{Planck} Collaboration V (2013) than for its X-ray selected
counterparts at lower $z$. \textit{Independent} evidence of HE prevalence at
high $z$ emerges from observations at multipoles $\ell\sim 3000$ of low
levels of integrated SZ effect (Reichardt et al. 2011, Efsthathiou \&
Migliaccio 2012, Lapi et al. 2012a, Hinshaw et al. 2013).

On the other hand, \emph{intermediate} types will occur even at $z \simeq
0.5$ when initial levels $k_c < 50$ keV cm$^2$ speed up the central cooling
relative to flattening of the outer slope $a$, and allow a number of strong
CCs to be already in place at $z\simeq 0.5$ (see Semler et al. 2012). In
turn, the flattening is speeded up in long-lived clusters, especially at low
$z$ and/or in poor environments where the accretion feeds on the wings of the
initial perturbation, while the ambient density is thinned out by the
accelerating cosmic expansion (see Lapi et al. 2012a). On the other hand,
cases with flat outer slope $a$ have been discerned in X-rays also at
intermediate $z$ by Eckert et al. (2013), Walker et al. (2013), by Sato et
al. (2014), and by McDonald et al. (2014). The feature has been discussed in
terms of super-clumps, that is, low entropy gas loosely bound in small, early
mergers. Alternatively, taking up from \S~2 one may think of it in terms of
still incomplete outskirts formation, when low DM masses $M(<R)$ drive slower
inflows with lower Mach numbers $\mathcal{M}^2<3$ that cause meager entropy
production.

In this complex context, where physical signals and prevailing processes are
still to be sorted out, we have stressed the value of the SZ effect in the
specific form of the \emph{patterns} and \emph{scaling} laws for $y(s)$ as
found in \S~3 and 4, separately for the cluster outskirts and cores. Here we
add that the l.o.s. integrations yielding $y(s)$ near the ends $s \simeq 0$
and $s \simeq R$ are dominantly contributed by outskirts or cores,
respectively. Thus such features of $y(s)$ can also probe \emph{intermediate}
states arising from asynchronous evolutions of cores and outskirts.

In fact, the relative timing of inner and outer ICM processes constitutes a
pressing physical issue to be directly addressed with SZ observations. The
patterns $y(s)$ can be pursued with developing ground instrumentation such as
\textsl{MUSTANG}, \textsl{BOLOCAM}, and \textsl{CARMA}, that have already
implemented sensitivities at sub-arcminute resolutions combined with a range
of arc-minute fields of view, appropriate for mapping individual features or
distant clusters in times from several to a few hours; see the discussions by
Mroczkowski et al. (2012) with their $S/N$ maps of MACS J0717.5+3745, by
Plagge et al. (2013), and by Mantz et al. (2014) on XLSSU J0217-0345 at
$z=1.9$, aiming at $S/N\sim 6$ in $3$ h for similar objects. We note that
comparatively shorter observations per cluster are required for LE/HE
classifications, while the independent core and outskirts patterns that we
stress in \S~4 and 6 allow a separate census and alleviate requirements on
field of view and resolution.

Such a strategy can be effective out to $z\sim 2$, where high-frequency SZ
signals increasingly mix with the far-IR emissions from violent, dusty star
formation in proto-spheroidal galaxies, often amplified by gravitational
lensing (Lapi et al. 2012b). In fact, such emissions provide an effective
probe for protoclusters, the first step toward cluster formation (see Davies
et al. 2014, Clements et al. 2014). Other contaminations by (possibly lensed)
background radiosources or by the CMB may be relevant at a single, low
frequency, but can  be cleaned out with standard procedures involving
multifrequency observations, as discussed by Melin et al. (2006), Hasler et
al. (2012), and Reichardt et al. (2013); see also Shirokoff et al. (2011).

Finally, we note that the $z$-depending physical conditions of the ICM
discussed here are particularly relevant when clusters are used to probe
cosmology in depth (see Vikhlinin et al. 2009). The use of reliable ICM
templates (different from self-similar) as we propose for LE or HE types will
help polish away the residual $10\%$ scatter that still hinders relating
strictly such homologous SZ and X-ray quantities as $Y_{500}$ and $Y_X$ (see
McCarthy et al. 2003; \textsl{Planck} Collaboration XVI, 2014). With
nonthermal pressure included as recalled in \S~5, they will also help in
pinning down cluster masses to closely constrain the values of cosmological
parameters.

\begin{acknowledgements}
We thank G. De Zotti, R. Fusco-Femiano, and K. Husband for helpful
discussions. We acknowledge the comments from our referee that stimulated us
to improve our presentation. Work supported in part by the MIUR PRIN
2010/2011 `The dark Universe and the cosmic evolution of baryons: from
current surveys to Euclid', and by the INAF PRIN 2012/2013 `Looking into the
dust-obscured phase of galaxy formation through cosmic zoom lenses in the
Herschel Astrophysical Terahertz Large Area Survey'. A.L. thanks SISSA for
the warm hospitality.
\end{acknowledgements}

\end{document}